\documentstyle[preprint,aps,prd]{revtex}
\begin{document}

\tighten

\title{Interplay of Non-Relativistic and Relativistic
Effects in Neutrino Oscillations}

\author{D. V. Ahluwalia
\footnotemark[1]
\footnotetext[1]{E--mail: av@p25hp.lanl.gov.} and T. Goldman
\footnotemark[2]
\footnotetext[2]
{E--mail: goldman@t5.lanl.gov.}}

\address{Los Alamos National Laboratory,
Los Alamos, NM 87545, USA}

\maketitle

\begin{abstract}

A theoretical structure that involves neutrino mass eigenstates
at non relativistic as well as relativistic energies is presented.
Using this framework, we find that if
the particle X, with mass $33.9$ 
MeV, of the KARMEN collaboration anomaly is identified
with the third neutrino mass  eigenstate, then the present limit
of $23$ MeV upper bound on the tau neutrino mass implies 
$\vert U_{\tau 3}\vert < 0.82$.

\end{abstract}

\pacs{PACS Number: 14.60.Pq}

\newpage

\section{Introduction}

The existing indications of neutrino oscillations \cite{Solar,LSND,Kamioka}
arise from 
data for neutrino energies in the sub-MeV to GeV range.
The neutrino oscillation phenomenology within the standard three--flavor
framework 
is developed on the basic
assumption that the neutrino mass
eigenstates that are superposed to yield the neutrino with flavor eigenstates
are relativistic.
This assumption is supported by the cosmological argument \cite{GB}  that the
sum of the neutrino masses have an upper bound of about $30\,\,\mbox{eV}$
\begin{equation}
\sum_\ell m(\nu_\ell)\,\le\, 30 \,\,\mbox{eV}\quad.\label{eq:con}
\end{equation}
However, very recent astronomical observations \cite{news1,pc1,news2}
raise potentially 
serious questions on the validity of the standard
cosmological model. First \cite{news1,pc1}, the UC Berkeley's
Extreme Ultraviolet Explorer  satellite's
observations of the Coma cluster of galaxies indicates that 
this cluster of galaxies may contain a submegakelvin cloud of $\sim 10^{13}
M_\odot$ of baryonic gas. Second \cite{news2}, the discovery by the
German X-ray satellite Rosat in which
a sample of 24 Seyfert galaxies contained 12 that were accompanied by a 
pairs of X--ray sources, almost certain to be high red shift quasars,
aligned on either side of the galaxy.
These observations and the associated interpretations, if correct, may 
place
severe questions to cosmological models that depend on the ratio
of photon to baryonic density in the
universe and its size.

We, therefore, consider it reasonable to relax the
cosmological constraint ({\ref{eq:con}) completely and explore
the resulting consequences for three-flavor neutrino oscillation 
framework.
The latest kinematic limits on neutrino masses 
are much less severe
\cite{taumass,PDB}:
\begin{eqnarray}
&&m(\nu_\tau) < 23 \,\,\mbox{MeV}\quad,\label{eq:tmass}\\
&&m(\nu_\mu) < 0.17 \,\,\mbox{MeV}\quad,\label{eq:mmass}\\
&&m(\nu_e) < 10-15\,\,\mbox{eV}\quad. \label{eq:emass}
\end{eqnarray}
The existing and the proposed neutrino oscillation experiments
involve neutrino energies from a fraction of a MeV
to several hundred  $\mbox{GeV}$
for the upper--energy end of the neutrino beams.
It remains possible that for some of the experiments
(or a sector of an experiment), the mass eigenstates 
$\vert\nu_2\rangle$ and/or $\vert\nu_3\rangle$
are non relativistic.
Towards the end of understanding such a possible situation
we now consider the 
interplay of non-relativistic and relativistic
mass eigenstates in three--flavor neutrino oscillation framework.

To establish the notational and conceptual context,
we begin with a brief review of the 
the standard three--flavor neutrino oscillation framework.

\section{Framework A: The Standard Three--Flavor Neutrino 
Oscillation Framework}

The standard three--flavor neutrino oscillation framework begins with the
assumptions that \cite{KPbook,BK,MPbook}

\begin{enumerate}

\item
There is no CP violation in the neutrino sector.

\item
Both the flavor and mass eigenstates are relativistic in the
laboratory frame.

\item
All neutrino mass eigenstates are stable.

\item
The weak flavor eigenstates 
$\vert\nu_e\rangle,\vert\nu_\mu\rangle,
\vert\nu_\tau\rangle$ are not identical to the mass eigenstates: 
\begin{equation}
\vert\nu_\ell\rangle
=\sum_\jmath U_{\ell\jmath}\,\vert\nu_\jmath\rangle
\quad.
\end{equation}
In the Maiani representation \cite{MR}, with CP phase $\delta$ set equal to zero,
the unitary neutrino--mixing matrix reads:
\begin{equation}
U(\theta,\,\beta,\,\psi)\,=\,
\left(
\begin{array}{ccccc}
c_\theta\,c_\beta &{\,\,}& s_\theta\,c_\beta &{\,\,}& s_\beta \\
-\,c_\theta\,s_\beta\,s_\psi\,-\,s_\theta\,c_\psi
&{\,\,}& c_\theta\, c_\psi\,-\,s_\theta\,s_\beta\,s_\psi
&{\,\,}& c_\beta\,s_\psi\\
-\,c_\theta\,s_\beta\,c_\psi\,+\,s_\theta\,s_\psi
&{\,\,}& -\,s_\theta\,s_\beta\,c_\psi\,-\,c_\theta\,s_\psi
&{\,\,}& c_\beta\,c_\psi
\end{array}\right),
\label{eq:umix}
\end{equation}
where $c_\theta\,=\,\cos(\theta)$, $s_\theta\,=\,\sin(\theta)$, etc.
The rows and columns are labeled as $U_{\ell \jmath}(\theta,\,\beta,\,\psi)$
with $\ell$ representing the neutrino--flavor eigenstates,
$\ell=e,\mu,\tau$, and $\jmath =1,2,3$ denoting the three mass eigenstates
with $m_1<m_2<m_3$.
\end{enumerate}

Within this framework, if a neutrino flavor state $\vert\nu_\ell\rangle$,
characterized by energy $E$, is created
at $x=0$ and $t=0$ then the probability that one observes
a neutrino state $\vert\nu_{\ell^\prime}\rangle$ at a neutrino--flavor
sensitive detector at $x=L\simeq t$ is \cite{DVA16} 
\begin{eqnarray}
&&P_{\ell\ell^\prime}\left(E,L,\{\eta_k\}\right)
= 
\delta_{\ell\,\ell'}
-\,\,4\,U_{\ell'\,1}\,U_{\ell\,1}\,U_{\ell'\,2}\,
U_{\ell\,2}\,\sin^2\left(\varphi^0_{2\,1}
\right)\nonumber\\
&&\quad\,-\,4\,U_{\ell'\,1}\,U_{\ell\,1}\,U_{\ell'\,3}\,
U_{\ell\,3}\,\sin^2\left(
\varphi^0_{3\,1}\right)
\,-\,4\,U_{\ell'\,2}\,U_{\ell\,2}\,U_{\ell'\,3}\,
U_{\ell\,3}\,\sin^2\left(
\varphi^0_{3\,2}\right)\, 
,\label{eq:d}
\end{eqnarray}
where the kinematic phase is defined as
\begin{equation}
\varphi^0_{\jmath\imath}
=
2\,\pi \,{L\over\lambda^{\rm osc}_{\jmath\imath }}\quad.\label{eq:kp}
\end{equation}
The flavor--oscillation length is defined as
\begin{equation}
\lambda^{{\rm {osc}}}_{\jmath
\imath}\,=\, {{2\,\pi}\over 1.27}\,{E\over{\Delta 
m^2_{\jmath\imath}}}\quad.\label{eq:ol}
\end{equation}
The five neutrino oscillation parameters $\{\eta_k\}$ that appear in 
Eq. (\ref{eq:d})
  are the two mass squared differences and
the three mixing angles: $\eta_1=\Delta m^2_{2 1}$,
$\eta_2=\Delta m^2_{3 2}$, $\eta_3=\theta$, $\eta_4=\beta$, 
and $\eta_5=\psi$. The third mass squared difference is then given by
$ \Delta m^2_{3 1}=\Delta m^2_{2 1}+\Delta m^2_{3 2}$.
Here, $1.27$ is the usual factor that
arises from expressing the neutrino energy
$E$ in $\mbox{MeV}\,\,(\mbox{or GeV})$, the source--detector distance
$L$ in meters $(\mbox{or kilometers})$, 
and the mass squared differences $\Delta m^2_{j\,k}$
in $\mbox{eV}^2\,$.

The kinematic phase  may also be written as: 
$\varphi^0_{\jmath\imath}
=1.27 \, \Delta m^2_{\jmath\imath } \times \left({L/ E}\right)$.
These kinematic phases may be modified for dynamical reasons 
\cite{DR1,AB,PRW}.

For all three pairs of
relativistic mass eigenstates, the relevant oscillation
lengths are given by
Eq. (\ref{eq:ol}). Again, because  
$ \Delta m^2_{3 1}=\Delta m^2_{2 1}+\Delta m^2_{3 2}$,
there are only two independent length scales: 
$\lambda^{\rm osc}_{2\,1}$ and
$\lambda^{\rm osc}_{3\,2}$. The third oscillation length is:
\begin{equation}
\lambda^{\rm osc}_{3\,1} = 
{{\lambda^{\rm osc}_{2\,1}\,\lambda^{\rm osc}_{3\,2}}
\over
{\lambda^{\rm osc}_{2\,1}+\lambda^{\rm osc}_{3\,2}}}\quad.
\end{equation}
The physical origin of this
lies in the assumption that all three mass eigenstates are
relativistic and that neutrino oscillations are inherently a
quantum mechanical interference phenomena.

Within the framework of the standard three--flavor neutrino oscillation
framework one cannot simultaneously
accommodate the known indications of the neutrino 
oscillations without questioning
a set of experiments or  the 
associated models (such as the standard solar model). 
One falls short of one length scale.
If one assumes that all experiments, and the relevant
theoretical understanding of the neutrino fluxes, are correct one is forced
to invoke 
additional physics, most commonly taken to be the existence of
a {\em sterile neutrino} \cite{sterile}. 

In principle,
there are six length scales in the three--flavor neutrino
oscillation phenomenology. Three are  associated
with the three mass eigenstates. The other three,  associated with
life times of the mass eigenstates, are 
rendered irrelevant
by the assumption of the stability of all three mass eigenstates.
In addition, matter effects and gravitational radii in astrophysical
environments contain additional length scales (see below
on how gravitational radii may come into the picture).

A theoretical structure that contains in it (a) 
the standard three--flavor neutrino oscillation
framework for the zenith angle dependent atmospheric neutrino
anomaly (Energies in the GeV range), and (b)
a three--flavor neutrino oscillation
framework with one or two non-relativistic mass eigenstates for
part of the MeV data, provides an alternative hypothesis
to the assumption of sterile neutrino.

For the sake of simplicity we shall maintain the assumption
that all three neutrino mass eigenstates are stable.

\section{
Framework B: Neutrino Oscillations with a Superposition of two Relativistic and
one Non-Relativistic Mass Eigenstate} 

Consider a physical situation where 
we have two relativistic, $\vert\nu_1\rangle$ and
$\vert\nu_2\rangle$,  neutrino mass eigenstates and one that is 
non--relativistic, $\vert\nu_3\rangle$.

At $t=0$, $x=0$, assume that a source  creates
a neutrino of flavor $\ell$
\begin{equation}
\vert\nu_\ell\rangle\,=\,U_{\ell 1}\,\vert\nu_1\rangle
\,+\,U_{\ell 2}\,\vert\nu_2\rangle
\,+\,U_{\ell 3}\,\vert\nu_3\rangle\quad.
\end{equation}
The spatial envelope, which is assumed to be ``sufficiently'' narrow, 
associated with $\{\vert\nu_1\rangle,\,\vert\nu_2\rangle\}$ evolves
towards the detector as $x\simeq t$, while the spatial envelope
of $\vert\nu_3\rangle$ evolves towards the detector as 
\begin{equation}
x\simeq (p/{m_3})\,t = \left({{\sqrt{2 m_3(E-m_3)}}\over
{m_3}}\right) t\quad.
\end{equation}
Therefore, the 
$\{\vert\nu_1\rangle,\,\vert\nu_2\rangle\}$ arrives at the 
detector at time $t_I\simeq L$, while the
$\vert\nu_3\rangle$ arrives at the detector at 
a later time $t_{II} \simeq m_3 L/\sqrt{2 m_3(E-m_3)}$. 
For the above considered energies and
for a source--detector distance of a few tens (or greater) of meters the 
detected $\vert\nu_{\ell^\prime}\rangle$ has no overlap with
$\vert\nu_3\rangle$ at the registered event at $t_I$, and similarly
the detected $\vert\nu_{\ell^\prime}\rangle$ has no overlap with
$\{\vert\nu_1\rangle,\,\vert\nu_2\rangle\}$ at the registered event at $t_{II}$.
With these observations at hand one can easily evaluate the modification
to the neutrino oscillation probability (\ref{eq:d}). The modified
expression reads:
\begin{eqnarray}
P_{\ell\ell^\prime}\left(E,L,\{\xi_k\}\right)
= 
\left(U_{\ell^\prime 3}\,U_{\ell 3}\right)^2 + && {\Big[} \left(
U_{\ell^\prime 1}\,U_{\ell 1}
\,+\,
U_{\ell^\prime 2}\,U_{\ell 2}\right)^2 \nonumber\\
&&-\,\,4\,U_{\ell'\,1}\,U_{\ell\,1}\,U_{\ell'\,2}\,
U_{\ell\,2}\,\sin^2\left(\varphi^0_{2\,1}\right){\Big]}
\quad.\label{eq:modd}
\end{eqnarray}
The first term on the rhs of the above equation
is the contribution
from the non--relativistic mass eigenstate
 to the 
$P_{\ell\ell^\prime}\left(E,L,\{\xi_k\}\right)$ at the $\nu_{\ell^\prime}$
event
at time $t_{II}$ while the second term is the
contribution  from the relativistic mass eigenstates
 to the  
$\nu_{\ell^\prime}$
event at the earlier time $t_I$.
Contained in the above expression is the 
the fundamental ``collapse of the wave packet'' postulate of the 
orthodox interpretation of the quantum mechanics.
That is, given a single $\nu_\ell$ emitted at the source, 
if the event occurs at $t_I$ no event occurs at $t_{II}$, and vice versa.

There are several observations that one may make about the result
(\ref{eq:modd}). These observations follow.

\begin{enumerate}

\item[{\bf I.}]

The neutrino oscillation probability now  contains only one length scale,
$\xi_1=\Delta m^2_{2 1}$, 
 $\xi_2=\theta$, $\xi_3=\beta$, 
and $\xi_4=\psi$.
However, this loss of a length scale is related to a manifestly
different expression for neutrino oscillation probabilities.

\item[{\bf II.}]

If $m_3$ is in the range 
of a fraction of an MeV to a few tens of MeV, one cannot base the analysis
of existing neutrino oscillation data in terms of Eq. (\ref{eq:d}), or
Eq. (\ref{eq:modd}), alone. 
For the zenith--angle dependence of the  atmospheric neutrino anomaly data
\cite{Kamioka}
neutrino energies in the GeV range
are involved. This meets the requirements under which
Eq. (\ref{eq:d}) is applicable.
On the other hand, the physical situation for part of the
LSND events (energy 
range between 20 MeV and Michel spectrum cutoff of 52.8 Mev \cite{LSND}),
and the reactor experiments (average $\overline{\nu}_e$ energy about 5 MeV 
\cite{R1,R2,R3,R4}), the physical conditions for
Eq. (\ref{eq:modd}) may be satisfied. 

\item[{\bf III.}]

For the solar--neutrino deficit it is conceivable 
that $m_3$ may be such that
an energy--dependent transition takes place from  the non-relativistic regime
to relativistic regime, thus accounting for the apparent energy dependence
of the solar--neutrino deficit
in an unconventional manner.
Consider the situation where the length
scales are such that all $\sin^2\left(\varphi^0_{\jmath\imath}\right)$
average to $1/2$. Then
\begin{equation}
P_{ee}\left[\mbox{Eq.}\,\,(\ref{eq:d})\right]
\,=\,1\,-\,2\,U^2_{e1}\,U^2_{e2}
\,-\,2\,U^2_{e1}\,U^2_{e3}
\,-\,2\,U^2_{e2}\,U^2_{e3}\quad,
\end{equation}
\begin{equation}
P_{ee}\left[\mbox{Eq.}\,\,(\ref{eq:modd})\right]
=U^4_{e1} \,+\,U^4_{e2} \,+\,U^4_{e3}\quad.
\end{equation}
Exploiting unitarity of the mixing matrix we immediately see that
above speculation, within the defined context, has no consequence because
unitarity implies
$
P_{ee}\left[\mbox{Eq.}\,\,(\ref{eq:d})\right] \,=\,
P_{ee}\left[\mbox{Eq.}\,\,(\ref{eq:modd})\right]\,.
$

\item[{\bf IV.}]

In astrophysical environments, such as for neutrinos observed in 
the supernova 1987a \cite{1987a,MRoos}, 
it may happen that $m_3$ is such that the state
$\vert \nu_3\rangle$
can not escape the astrophysical environment.
That is, $p/m$ associated with the 
$\vert \nu_3\rangle$ is less than the escape velocity
\begin{equation}
\mbox{Non Relativistic:}\quad E \,<\, m\,+\,{{r_g}\over {2\, r}}\quad,
\end{equation}
where $r_g\equiv 2\, G M$ is the gravitational radius 
of the astrophysical object.
Under these
circumstances Eq. (\ref{eq:modd}), for a detector at Earth, reduces to:
\begin{eqnarray}
&&P_{\ell\ell^\prime}\left(E,L,\{\xi_k\}\right)
= 
\left(U_{\ell^\prime 3}\,U_{\ell 3}\right)^2 
\,\Theta\left(E \,-\, m\,+\,{{r_g}\over {2\, r}}\right)\nonumber\\
&&\,+\,
\left(
U_{\ell^\prime 1}\,U_{\ell 1}
\,+\,
U_{\ell^\prime 2}\,U_{\ell 2}\right)^2 
-\,\,4\,U_{\ell'\,1}\,U_{\ell\,1}\,U_{\ell'\,2}\,
U_{\ell\,2}\,\sin^2\left(\varphi^0_{2\,1}\right)
\,.\label{eq:moddd}
\end{eqnarray}
In the above expression
$\Theta(\cdots)$ is the Heaviside function, vanishing for its argument
less than zero and equal to unity for its argument greater or equal
to zero. Note that depending upon how much the argument of 
$\Theta(\cdots)$ exceeds zero, the arrival of the $\vert\nu_3\rangle$
could be enormously delayed. 
Further, if  $U_{\ell'3}$ is small, the delayed signal
could easily also be too small to be detected.
For example, in the Kamiokande detector, an oscillation amplitude $\lesssim 0.5$
would have led, within statistical fluctuations, to
as few as one additional event beyond the
time windows 
identified with the 1987A supernova pulse (and so could have have failed to be
recognized). This possibility was ignored earlier only due to the cosmological
constraints which are now in question.
\end{enumerate}

Now we rewrite Eq. (\ref{eq:modd}) in a form that makes the deviations
of result (\ref{eq:modd}) explicit from the corresponding two--flavor scenario
with two relativistic mass eigenstates.  This new form of
Eq. (\ref{eq:modd}) reads:
\begin{equation}
P_{\ell\ell^\prime}\left(E,L,\{\xi_k\}\right)
= 
\Delta_{\ell\ell^\prime}
\,-\,A_{\ell\ell^\prime}\,\sin^2\left(\varphi^0_{2\,1}\right)
\quad,\label{eq:nf}
\end{equation}
with
$ \Delta_{\ell\ell^\prime}\,=\,
\left(U_{\ell^\prime 3}\,U_{\ell 3}\right)^2 \,+\,
\left(
U_{\ell^\prime 1}\,U_{\ell 1}
\,+\,
U_{\ell^\prime 2}\,U_{\ell 2}\right)^2 $, or more explicitly
\begin{equation}
\Delta=
\openone +
\left(
\begin{array}{ccccc}
\mbox{s}^4_\beta+\mbox{c}^4_\beta -1 &{\,\,}&
2\mbox{c}^2_\beta \mbox{s}^2_\beta \mbox{s}^2_\psi &{\,\,}&
2\mbox{c}^2_\beta \mbox{s}^2_\beta \mbox{c}^2_\psi \\
2\mbox{c}^2_\beta \mbox{s}^2_\beta \mbox{s}^2_\psi &{\,\,}&
2 \mbox{c}^2_\beta \mbox{s}^2_\psi(\mbox{c}^2_\beta \mbox{s}^2_\psi-1) &{\,\,}&
2\mbox{c}^4_\beta \mbox{c}^2_\psi\mbox{s}^2_\psi \\
2\mbox{c}^2_\beta \mbox{s}^2_\beta \mbox{c}^2_\psi &{\,\,}&
2\mbox{c}^4_\beta \mbox{c}^2_\psi\mbox{s}^2_\psi&{\,\,}&
2 \mbox{c}^2_\beta \mbox{c}^2_\psi(\mbox{c}^2_\beta \mbox{c}^2_\psi-1)
\end{array}
\right)\,,\label{eq:delta}
\end{equation} 
and
\begin{equation}
A_{\ell\ell^\prime}\equiv 4\,U_{\ell'\,1}\,U_{\ell\,1}\,U_{\ell'\,2}\,
U_{\ell\,2}
\quad,
\end{equation} 
where  $\openone$ is a $3\times 3$ identity matrix.

Only when both $\beta$ and $\psi$ vanish does Eq. (\ref{eq:nf})
reduce to the expression for
the two flavor scenario
with two relativistic mass eigenstates, for then 
$U(\theta,\,\beta,\,\psi)$ becomes block diagonal with no mixing with 
$\vert \nu_3\rangle$, $\Delta (\beta=\psi=0) =\openone$, and
\begin{equation}
A =
\left(
\begin{array}{ccc}
\mbox{s}^2_{2\theta} & -\,\mbox{s}^2_{2\theta} & 0 \\
-\,\mbox{s}^2_{2\theta} & \mbox{s}^2_{2\theta} & 0\\
0 & 0 & 0
\end{array}
\right)\quad.
\end{equation}

\section{Framework C:
Neutrino Oscillations with a Superposition of one Relativistic and
two Non-Relativistic Mass Eigenstates} 

For completeness, we now
consider a physical situation where 
we have one relativistic, $\vert\nu_1\rangle$, and two
non--relativistic, $\vert\nu_2\rangle$ and
 $\vert\nu_3\rangle$,  neutrino mass eigenstates. The configuration of
masses is so chosen that $m_1\ll m_2 \le m_3$, and
\begin{equation}
\frac{\delta m_{ 3 2}}{\langle m \rangle}\ll 1\,;\quad
\delta m_{ 3 2} \equiv m_3-m_2\,,\quad
 \langle m \rangle\equiv
(m_3+m_2)/2\quad.
\end{equation}

Then, arguments similar to those above yield:
\begin{eqnarray}
P_{\ell\ell^\prime}\left(E,L,\{\xi^\prime_k\}\right)
= 
\left(U_{\ell^\prime 1}\,U_{\ell 1}\right)^2 + && {\Big[} \left(
U_{\ell^\prime 2}\,U_{\ell 2}
\,+\,
U_{\ell^\prime 3}\,U_{\ell 3}\right)^2 \nonumber\\
&&-\,\,4\,U_{\ell'\,1}\,U_{\ell\,1}\,U_{\ell'\,2}\,
U_{\ell\,2}\,\sin^2\left(\zeta^0_{3\,2}\right){\Big]}
\quad.\label{eq:modddd}
\end{eqnarray}
In Eq. (\ref{eq:modddd}),
the first term is now the contribution from the relativistic mass
eigenstate, while the second interference term  arises from the
two non--relativistic mass eigenstates.

The new kinematic phase  $\zeta^0_{3\,2}$
that appears in Eq. (\ref{eq:modddd})
is defined as\footnote{Here, 
we keep $\hbar$ and $c$ explicit in order
that appropriate counterparts of $1.27$ may be
introduced as dictated by the magnitude of relevant variables.}
\begin{equation}
\zeta^0_{3 2}=
\frac
{c^3}
{2\,\hbar}\,
\frac
{\delta m_{3 2}\,\langle m\rangle\,L}
{\left[2 \langle m \rangle c^2 \left(E- \langle m\rangle c^2\right)
\right]^{1/2}}
\quad.\label{eq:nkp}
\end{equation}
To compare the 
new kinematic phase that now enters our considerations
with Eqs. (\ref{eq:kp}) and (\ref{eq:ol}),
we rewrite Eq. (\ref{eq:nkp}) as:
\begin{equation}
\zeta^0_{3 2}
=
2\,\pi \,{L\over\Lambda^{\rm osc}_{ 3 2}}\quad,\label{eq:np}
\end{equation}
with the  new flavor--oscillation length given by
\begin{equation}
\Lambda^{{\rm {osc}}}_{3 2}\,=\, 
=\frac{ 4 \sqrt{2} \pi\hbar}
{\delta m_{3 2}\, c}
\left(
\frac{E-\langle m \rangle c^2}
{\langle m \rangle c^2}
\right)^{1/2}\quad.\label{eq:nosl}
\end{equation}
Denoting 
$\left[2 \langle m \rangle  \left(E- \langle m\rangle c^2\right)
\right]^{1/2}$ by $\langle p \rangle$, Eqs. (\ref{eq:nkp}) and
(\ref{eq:nosl}) take the 
simpler form
\begin{equation}
\zeta^0_{3 2}=
\frac
{c^2}
{4\,\hbar}\,
\frac
{\Delta m^2_{3 2} \, L}
{\langle p \rangle}\quad,
\end{equation}
and
\begin{equation}
\Lambda^{{\rm {osc}}}_{3 2}\,=\,
\frac{8\pi\hbar}
{c^2}\,
\frac{\langle p \rangle}{\Delta m^2_{3 2}}
\quad.
\end{equation}
However, this rewriting should not leave the impression
that there is now no $m$--dependence.
Unlike the standard ultra--relativistic 
case, where the $m$--dependence
of $E$ disappears in $E\simeq 
p c$,  we now the have an explicit $m$--dependence via
$\langle p \rangle=
\left[2 \langle m \rangle  \left(E- \langle m\rangle c^2\right)
\right]^{1/2}$.

The neutrino--flavor oscillation probability now depends on the 
differences in masses of the non-relativistic mass
eigenstates, is independent of the mass of the lowest lying mass eigenstate,
and has an explicit dependence on the absolute scale of masses of
non-relativistic mass eigenstates via $\langle m \rangle$.

The five neutrino oscillation parameters are:
$\xi_1^\prime=\delta m_{3 2}$, 
$\xi_2^\prime=\langle m \rangle$, 
$\xi_3^\prime=\theta$, 
$\xi_4^\prime=\beta$, 
and $\xi_5^\prime=\psi$.

In astrophysical contexts, Eq. (\ref{eq:modddd}) is modified 
similarly to before (cf., Eq.{\ref{eq:moddd})
\begin{eqnarray}
P_{\ell\ell^\prime}&&\left(E,L,\{\xi^\prime_k\}\right)
= 
\left(U_{\ell^\prime 1}\,U_{\ell 1}\right)^2 + 
\Theta\left(E \,-\, \langle m\rangle\,+\,{{r_g}\over {2\, r}}\right)\nonumber\\
&& \times{\Big[} \left(
U_{\ell^\prime 2}\,U_{\ell 2}
\,+\,
U_{\ell^\prime 3}\,U_{\ell 3}\right)^2 
-\,\,4\,U_{\ell'\,1}\,U_{\ell\,1}\,U_{\ell'\,2}\,
U_{\ell\,2}\,\sin^2\left(\zeta^0_{3\,2}\right){\Big]}
\quad.\label{eq:astroC}
\end{eqnarray}
It is again transparent that a superposition of
relativistic and  non-relativistic mass eigenstates has a 
a rich structure.  
Important physical information can be gathered about neutrino
properties
by carefully analyzing the existing data and 
designing future neutrino experiments at all feasible energies.
If the physically realized scenario occurs in a set of experiments that
corresponds to
Frameworks $\mbox{A}+\mbox{B}$, the mass--parameters that enter the 
neutrino oscillation phenomenology evolves from dependence
on  
$\Delta m^2_{2 1}$ to dependence on $\Delta m^2_{2 1}$ and $\Delta m^2_{3 2}$. 
If the physically realized scenario occurs in a set of experiments that
corresponds to
Frameworks $\mbox{A}+\mbox{C}$, the mass--parameters that enter the 
neutrino oscillation phenomenology change from explicit dependence on
$m_2$ and $m_3$ to 
$\Delta m^2_{2 1}$ and $\Delta m^2_{3 2}$.

\section{KARMEN Anomaly: A Hint of a Non--Relativistic Mass Eigenstate ?}

KARMEN continues to see an interesting anomaly in the time distribution of
$\nu_e$ and $\overline{\nu}_\mu$
\cite{Kanomaly,pc2}.
Tentatively, the KARMEN collaboration
interprets this anomaly as an indication of a new neutral particle with its mass
and lifetime (electromagnetic decay) as follows:
\footnote{This section evolved in discussions with Dr. Hywel White.
We wish to acknowledge his physics contributions, and thank him
warmly for sharing his wisdom and thoughts with us.}
\begin{equation}
\mbox{KARMEN Particle X:}\quad m_x=33.9\,\,\mbox{MeV}\,,\quad
\tau_x>0.3\,\,\mu\mbox{s}\quad.
\end{equation}
We consider the identification of it with $\vert\nu_3\rangle$, so that
\begin{equation}
m_x\,=\,m_3\quad.\label{eq:mx3}
\end{equation}

Within the three flavor neutrino oscillation framework,
the expectation value of the mass operator
in the state 
$
\vert\nu_\tau\rangle\,=\,\sum_\jmath U_{\tau \jmath}\,\vert\nu_\jmath\rangle
\, is
$
\begin{equation}
m({\nu_\tau}) \,=\, U^2_{\tau 1}\, m_1 + U^2_{\tau 2}\, m_2 +
U^2_{\tau 3}\, m_3\quad.
\end{equation}
Guided by the current limits on 
the remaining neutrino masses given by Eqs. (\ref{eq:mmass}) and 
(\ref{eq:emass})
we assume 
\begin{equation}
U^2_{\tau 1}\, m_1 + U^2_{\tau 2} \,m_2\, \ll\, 
U^2_{\tau 3} \,m_3 \quad.
\end{equation}
Using the identification
(\ref{eq:mx3})
we obtain
\begin{equation}
\vert U_{\tau 3}\vert \,<\,0.82\quad.
\end{equation}

To exploit the phenomenological consequences of our proposal to
identify the KARMEN anomaly with the third mass eigenstate one must
now incorporate the life time of $\vert \nu_3\rangle$ in the 
above proposed framework. In the context of supernova 1987a such a 
formalism can be found in the work of Frieman, Haber, and Freese \cite{FHF}.

In reference to the PSI's ``search for a neutral particle of mass
33.9 MeV in pion decay'' \cite{PSI}
\begin{equation}
\pi^+\,\rightarrow\, \mu^+ \,+\,X\quad;\label{eq:x}
\end{equation}
if $\vert X\rangle$ is indeed $\vert\nu_3\rangle$, the kinematics of
the KARMEN anomaly  may be significantly different
from the kinematics associated with the PSI--considered decay
(\ref{eq:x}). While the 
details of this difference have not been worked out in detail, 
it is clear that identifying
the KARMEN anomaly with the third mass eigenstate means that the
particle X will be found not only in the decay
(\ref{eq:x}) but also (provided the appropriate energy threshold requirements
are met) in the tau, muon, and electron neutrino beams. 
\footnote{
The PSI experiment may be considered as a search for the KARMEN X in the 
$\nu_\mu$ beam.}
Above the production threshold, 
the  relative X--event rates for these beams
are expected to be roughly in the ratio 
\begin{equation}
U_{\tau 3}^2: U_{\mu 3}^2
:U_{e 3}^2\quad,
\end{equation}
for equal fluxes of $\nu_\tau,\,\,\nu_\mu,\,\,\nu_e$ respectively.

\section{Concluding Remarks}

In this paper we have established that  interplay of
non-relativistic and relativistic
effects in neutrino oscillations may play  an important role
in understanding the existing indications of neutrino oscillations
and design of future experiments. 

As  an alternate hypothesis to the assumption of sterile neutrino we
have put forward a 
theoretical structure that contains in it the
(a) the standard three--flavor neutrino oscillation
framework for the GeV 
data on the zenith angle dependent atmospheric neutrino
anomaly, and 
(b) a three--flavor neutrino oscillation
framework with one or two non-relativistic mass eigenstates for
part of the MeV data 
on neutrino oscillation experiments.

If the particle X, with mass $33.9$ 
MeV, of the KARMEN collaboration anomaly is identified
with the third neutrino mass  eigenstate, then the present limit
of $23$ MeV upper bound on the tau neutrino mass leads to $\vert U_{\tau 3}
\vert < 0.82$.

Our brief discussion on KARMEN anomaly 
suggests that one may not only need to relax the assumption on
the relativistic nature of all three mass eigenstates, as done
in the present paper, but in addition the stability of the neutrino mass
eigenstates may no longer be a viable hypothesis.

\end{document}